\documentclass[letterpaper]{mandc2019}
\usepackage{tabls}
\usepackage{cites}
\usepackage{epsf}
\usepackage{appendix,amsmath}
\usepackage{amsfonts}
\usepackage{ragged2e}
\usepackage[top=1in, bottom=1.in, left=1.in, right=1.in]{geometry}
\usepackage{enumitem}
\setlist[itemize]{leftmargin=*}
\usepackage{caption}
\title{Acceleration of Source Iteration using the Dynamic Mode Decomposition}
\author{%
  \textbf{Ryan G.\ McClarren$^1$, Terry S. Haut$^2$}
  \\ \\
  $^1$Dept. of Aerospace and Mechanical Engineering  \\
  University of Notre Dame \\
  Notre Dame, Indiana 46556 \\ 
\\
  $^2$ Lawrence Livermore National Laboratory  \\ 
    Livermore, California \\ 
\\
  \tt{rmcclarr@nd.edu}, \tt{haut3@llnl.gov}
}
\newcommand{\authorHead}      
           {First author}  
\newcommand{\shortTitle}      
           {Paper Title }  
\begin{document}
\maketitle
\justify 

\begin{abstract}
We present a novel acceleration technique for improving the convergence of source iteration for discrete ordinates transport calculations. Our approach uses the idea of the dynamic mode decomposition (DMD) to estimate the slowly decaying modes from source iteration and remove them from the solution.  The memory cost of our acceleration technique is that the scalar flux for a number of iterations must be stored; the computational cost is a singular value decomposition of a matrix comprised of those stored scalar fluxes.  On 1-D slab geometry problems we observe an order of magnitude reduction in the number of transport sweeps required compared to source iteration and that the number of sweeps required is independent of the scattering ratio in the problem.  These observations hold for an extremely heterogeneous problem and a 2-D problem.  In 2-D we do observe that the effectiveness of the approach slowly degrades as the mesh is refined, but is  still about one order of magnitude faster than source iteration.
\end{abstract}

\section{INTRODUCTION} 
 In scientific computing we are used to taking a known operator and making approximations to it, this is the basis for most numerical methods. Conversely, without knowledge of the operator, it is possible to  use just the action of the operator to generate approximations to it. This is done in Krylov methods where the action of a linear operator is used to build a subspace to find solutions in \cite{Warsa:2004p1085}. One can think of such approaches as data-driven methods.
 
 One data driven method is the dynamic mode decomposition (DMD) \cite{Schmid:2010ee,Schmid:2010hh,Schmid:2011ec} which uses the action of the operator to infer eigenmodes of an operator.  DMD has enjoyed success in the fluid dynamics community as a way to compare simulation and experiment. This is possible because measured data can also be used in DMD, and the DMD modes can be directly compared between experiment and simulation.
 DMD has also been recently shown to be capable of computing time eigenvalues of neutron transport problems by computing the evolution of the system over time \cite{mcclarren2018calculating}. This approach works for sub-critical as well as super-critical systems and finds the eigenmodes that are significant in a particular system.
 
 For the purpose of this paper we are using DMD to infer information about the source iteration operator in a discrete ordinates (S$_N$) transport calculation.  Using DMD we can estimate the slowest decaying modes in the iterative procedure and extrapolate to find an estimated converged solution. We begin with a description of DMD

\section{DYNAMIC MODE DECOMPOSITION} 
\label{sec:DMD}
Here we present the basics of the dynamic mode decomposition.  For more detail see \cite{Schmid:2010ee,Schmid:2010hh,Schmid:2011ec,Tu:2013dj}.

Consider a sequence of vectors $\{y_0, y_1, \dots, y_K\}$ where $y_k \in \mathbb{R}^N$. The vectors are related by a potentially unknown linear operator of size $N \times N$, $A$, as
  \[ y_{k+1} = A y_k.\]
  If we construct the $N \times K$ {\em data matrices} $Y_+$ and $Y_-$, 
  \[ Y_+ = \begin{pmatrix}   | & | & & | \\ y_1 & y_2 & \dots &y_K \\  | & | & & |  \end{pmatrix}
  \qquad Y_- = \begin{pmatrix}   | & | & & | \\ y_0 & y_1 & \dots &y_{K-1} \\  | & | & & |  \end{pmatrix}\]
  we can write \[ Y_+ = A Y_-.\]
  At this point we only need to know the data vectors $y_k$, they could come from a calculation, measurement, etc.
  As $K \rightarrow \infty$ we could hope to infer properties about $A$. 
  
We take the thin singular value decomposition (SVD) of $Y_-$ to write
  \[ Y_- = U \Sigma V^\mathrm{T},\]
  where $U$ is a $N \times K$ orthogonal matrix, $\Sigma$ is a diagonal $K\times K$ matrix with non-negative entries on the diagonal, and $V$ is a $K \times K$ orthogonal matrix.
   The SVD requires $O(NK^2)$ operations to compute.
   Later, we will want $K \ll N$, if, for example, $N$ is the number of unknowns in a transport calculation.
   Also, if the column rank of $Y_- < K$, then there is a further reduction in the SVD size.
   The matrix $U$ has columns that forms an orthonormal basis for the row space of $Y_-  \subset \mathbb{R}^N$. 
   Using the SVD we get 
   \[ Y_+ = A U \Sigma V^\mathrm{T}.\]
    If there are only $r < K$ non-zero singular values in $\Sigma$, we use the compact SVD where $U$ is $N \times r$, $\Sigma$ is $r \times r$, and $V$ is $K \times K$.
  
   We can rearrange the relationship between $Y_+$ and $Y_-$ to be 
  \[ Y_+ = A U \Sigma V^\mathrm{T} \qquad  \rightarrow \qquad U^\mathrm{T} A U =  U^\mathrm{T} Y_+ V \Sigma^{-1}.\]
   Define $\tilde{A} = U^\mathrm{T} A U = U^\mathrm{T} Y_+ V \Sigma^{-1}.$ This is a rank $K$ approximation to $A$.
   Using the approximate operator $\tilde{A}$, we can now find out information about $A$.
   The eigenvalues/vectors of $\tilde{A}$, \[ \tilde{A} w = \lambda w,\]
  are used to define the dynamic modes of $A$:
  \[ \varphi = \frac{1}{\lambda} U^\mathrm{T} Y_+ V \Sigma^{-1} w.\]
   The dynamic mode decomposition (DMD) of the data matrix $Y_+$ is then the decomposition of into vectors $\varphi$. The mode with the largest norm of $\lambda$ is said to be the dominant mode.

\section{DMD AND SOURCE ITERATION}
\label{sec:SI}
The discrete ordinates method for transport is typically solved using source iteration (Richardson iteration) and diffusion-based preconditioning/acceleration.
 Source iterations converge quickly for problems with a small amount of particle scattering.
 For strongly scattering media, the transport operator has a near nullspace that can be handled using a diffusion preconditioner.
 However, the question of efficiently preconditioning/accelerating transport calculation on high-order meshes with discontinuous fine elements is an open area of research.
 The approximate operator found from DMD can be used to remove this same near nullspace and improve iterative convergence {\em without the need for a separate preconditioner or diffusion discretization/solve}.

 The steady, single group transport equation with isotropic scattering can be written as 
\begin{equation}\label{eq:transport}
    L \psi = \frac{c}{4\pi} \phi + \frac{Q}{4\pi},
\end{equation}

where $c$ is the scattering ratio, $Q$ is a prescribed, isotropic source, and the streaming and removal operator is
\[ L = \left( \Omega \cdot \nabla + 1\right).\]
In this equation the angular flux is 
 $\psi(\mathbf{x},\Omega)$, with the direction-of-flight variable written $\Omega \in \mathbb{S}_2$, (i.e., $\Omega$ is a point on the unit sphere). The scalar flux is the integral of the angular flux over the unit sphere: 
 \[\phi(\mathbf{x}) = \int_{4\pi} \psi\,d\Omega  = \langle \psi \rangle.\]
 Source iteration solves the problem in Eq.~\eqref{eq:transport} using the iteration strategy
\begin{equation}\label{eq:SI}
\phi^{\ell} = \left\langle L^{-1} \left(\frac{c}{4\pi} \phi^{\ell-1} + \frac{Q}{4\pi}\right)\right\rangle,
\end{equation}
where $\ell$ is an iteration index.
 One iteration is often called a ``transport sweep".
 A benefit of source iteration is that the angular flux, $\psi$ does not have to be stored.
 As $c\rightarrow 1$,  the convergence of source iteration can be arbitrarily slow \cite{Adams:2001vt}.

 Rearranging the transport equation we see that source iteration is an iterative procedure for solving
\begin{equation}\label{eq:update}
     \phi - \left\langle L^{-1} \frac{c}{4\pi} \phi \right \rangle = \langle L^{-1} Q \rangle.
\end{equation}
We can define an operator $A$ and a vector $b$ to write Eq.~\eqref{eq:update} as
\[ (I-A)\phi=b.\]
 Therefore, the source iteration vectors are
\[\phi^{\ell + 1} = A \phi^\ell + b,\]
or, by subtracting successive iterations,
\[ \phi^{\ell + 1} - \phi^\ell = A (\phi^\ell - \phi^{\ell-1}).\]
 Therefore, we can cast the difference between iterates in a form that is amenable to the approximation of $A$ using DMD, $Y_+ = A Y_-,$
\[Y_+ = \left[ \phi^2 - \phi^1, \phi^3 - \phi^2, \dots, \phi^K - \phi^{K-1}\right],\]\[ Y_- = \left[ \phi^1 - \phi^0, \phi^2 - \phi^3, \dots, \phi^{K-1} - \phi^{K-2}\right].\]

 As before we define an approximate $A$ as the $K\times K$ matrix:
\[\tilde{A} = U^\mathrm{T} A U = U^\mathrm{T} Y_+ V \Sigma^{-1},\]
 We can use $\tilde{A}$ to construct the operator $(I-\tilde{A})^{-1}$ and use this to approximate the solution:
\begin{align*}
(I-A) (\phi - \phi^{K-1}) &= b - (I-A)\phi^{K-1}\\ 
&= b - \phi^{K-1} + (\phi^{K}-b)\\ 
&= \phi^K - \phi^{K-1}.
\end{align*}
 The difference $\phi - \phi^{K-1}$ is the difference between step $K-1$ and the converged answer.  We define a new vector $\Delta y$ as the length $K$ vector that satisfies
\begin{equation}\label{eq:dely}
\phi - \phi^{K-1} = U \Delta y.
\end{equation}
 We then substitute  and multiply by $U^\mathrm{T}$ to get 
\begin{equation}\label{eq:final_1}
(I - \tilde{A})\Delta y = U^\mathrm{T}(\phi^K - \phi^{K-1}).
\end{equation}
This is a linear system of size $K$ that we can solve to get $\Delta y$ and then compute the update to $\phi^{K-1}$ as
\begin{equation}\label{eq:finalDMD}
\phi \approx \phi^{K-1} + U \Delta y.
\end{equation}

 The algorithm is as follows
\begin{enumerate}
\item  Perform $R$ source iterations: $\phi^\ell = A \phi^{\ell-1} + b$.
 \item Compute $K$ source iterations to form $Y_+$ and $Y_-$. The last column of $Y_-$ we call $\phi^{K-1}$.
 \item Compute $\phi = \phi^{K-1} + U \Delta y$ as above.
\end{enumerate}
 Each pass of the algorithm requires $R+K$ source iterations.
 The $R$ source iterations are used to correct any errors caused by the approximation of $A$ using the SVD.
 It is easiest to assess convergence between the source iterations.
 This works regardless of the spatial discretization used.
 Other algorithms are possible:
\begin{itemize}
    \item Rather than extrapolate to an infinite number of iterations, we can use $\tilde{A}$ to approximate a finite number of source iterations.
    \item  We could use a coarsened vector $\bar{\phi}$  in the DMD procedure to reduce the memory/computational cost.
    \item We could use DMD in the low-order solve of a transport synthetic acceleration scheme.
    \item The DMD acceleration could be wrapped in a Krylov solver \cite{Warsa:2004p1085} to further improve performance.
\end{itemize}

\section{NUMERICAL RESULTS}

\subsection{Slab Geometry Examples} 
\label{sec:slab}
\begin{figure}[!htb]
  \centering
  \includegraphics[width=\textwidth]{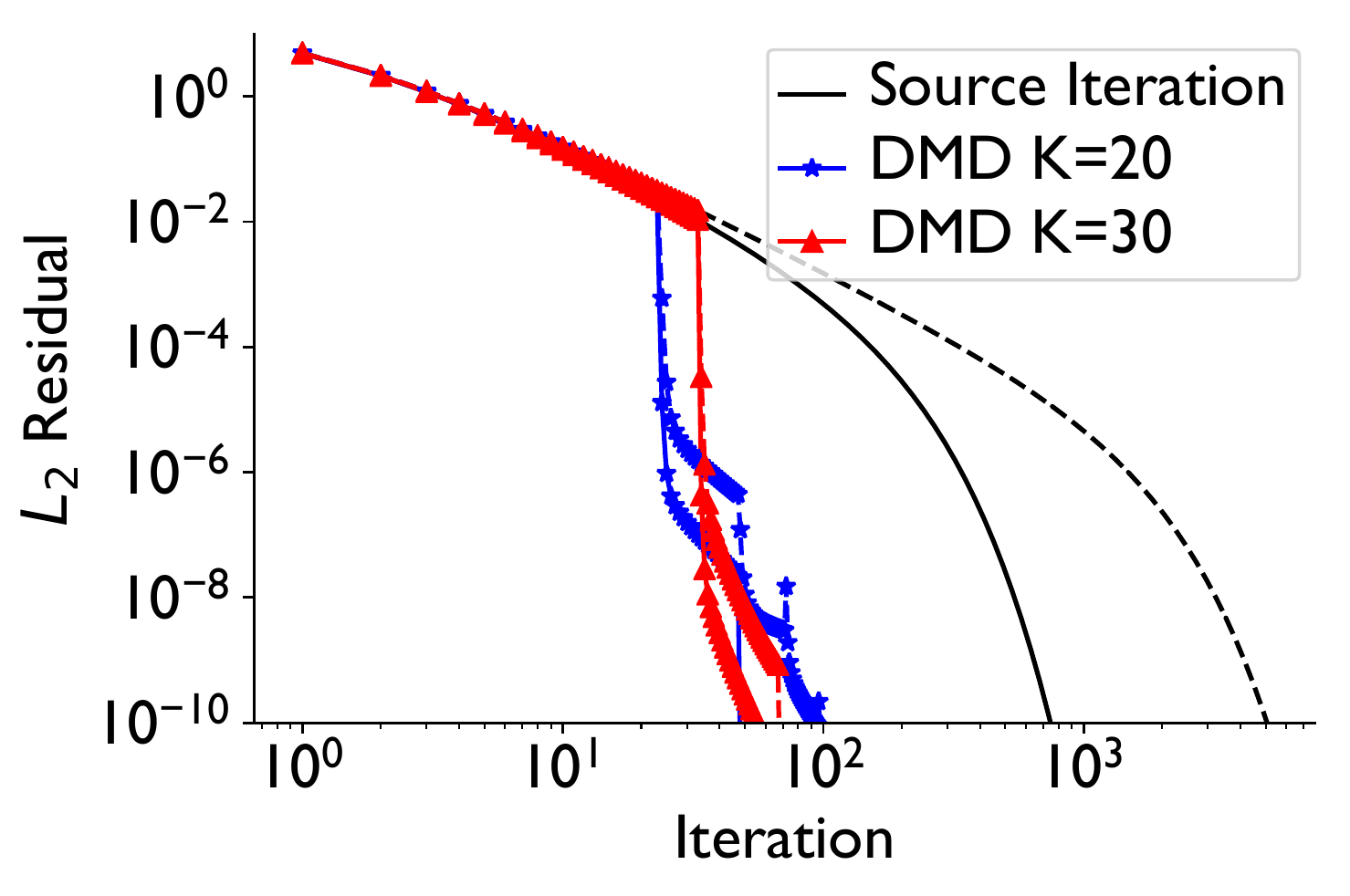}
  \caption{Residual for the homogeneous slab geometry problem. The horizontal axis is the number of transport sweeps.  The residual does not change during the steps computed to estimate the DMD update.}   
  \label{fig:residual_slab}
\end{figure}

{DMD works perfectly on a homogeneous slab, the ur-demonstration problem for acceleration schemes.}
We consider a slab 50 mean-free paths thick with vacuum boundaries and a scattering ratio of $c=0.99$ and $1.0$ and 400 spatial zones, $S_8$ angular discretization, and the diamond difference spatial discretization. The results from source iteration and DMD with $R=4$ and different values of $K$ are shown in Figure \ref{fig:residual_slab}. In this figure solid lines are $c=0.99$ results and dashed lines are $c=1.0$ From the figure we see that the DMD results converge about one order of magnitude fewer transport sweeps than source iteration. In the figure, we can see that between DMD updates, the convergence follows source iteration's trend as the solutions to estimate $\tilde{A}$ are computed.

To further explore the behavior of the DMD acceleration we solve a homogeneous slab problem with 1000 cells and 50 mean-free paths in the slab for various scattering ratios. Table \ref{table:slab} shows the number of transport sweeps required to solve this problem as a function of the scattering ratio and the number of sweeps used in the DMD update. It does appear that there is an optimal value for $K$, though we have observed this to be somewhat problem dependent. Regardless of the value of $K$ chosen, the number of iterations required appears to be independent of the scattering ratio for DMD.

\begin{table}[!htb]
  \centering
  \caption{\bf Number of iterations (transport sweeps) for the homogeneous slab geometry problem. The DMD results used $R=4$.}
  \label{table:slab} 
  \begin{tabular}{r|rrrrrrrr}
\hline
  $K/c$ &  0.1 &  0.5 &  0.9 &  0.99 &   0.999 &   0.9999 &   0.99999 &   0.999999 \\
\hline
  3 &  8 & 15 & 39 & 70 &  70 &  70 &  70 &  70 \\
  5 & 10 & 11 & 28 & 90 &  90 &  90 &  90 &  90 \\
 10 & 15 & 15 & 29 & 60 & 140 & 140 & 140 & 140 \\
 20 & 25 & 25 & 25 & 49 &  74 &  76 &  76 &  76 \\
 50 & 55 & 55 & 55 & 56 &  57 &  57 &  57 &  57 \\
 SI & 6 & 17 & 89 & 637 & 2439 & 3681 & 3889 & 3911 \\
\hline
\end{tabular}
\end{table}

\begin{figure}[!htb]
  \centering
  \includegraphics[width=\textwidth]{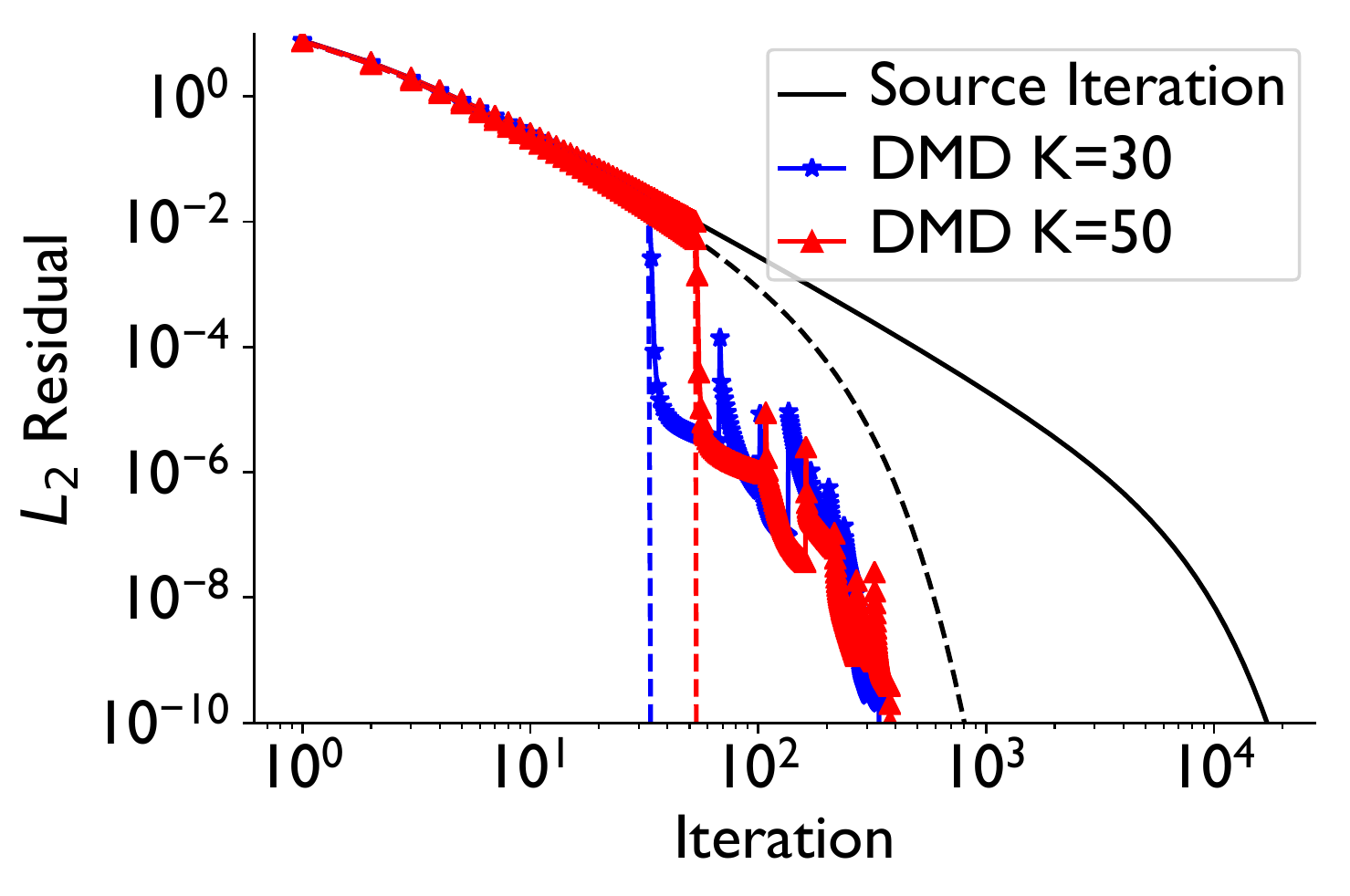}
  \caption{Residual for heterogeneous slab geometry problem with $c=0.9999$. Two cases are shown $p=5$ (dashed) and $p = 8$ (solid).}   
  \label{fig:residual_slab_het}
\end{figure}

We have observed that performance does degrade on an {\em ad absurdum} heterogeneous problem inspired by \cite{Warsa:2004p1085}.
 To demonstrate this, we consider a problem with vacuum boundaries, 1000 cells, unit domain length, with $c = 0.9999$ and \[\sigma_\mathrm{t} = \begin{cases} 2^p & \text{cell number odd} 
\\ 2^{-p} & \text{cell number even}\end{cases}.\]
 In Figure \ref{fig:residual_slab_het} we see convergence for $p=5$ (dashed) and $p = 8$ (solid), a factor of about 1000 and  $6.5\times 10^4$ between thick and thin cells, respectively. In this problem more iterations are necessary, however, there is still about and order of magnitude fewer transport sweeps for the DMD accelerated calculations. The results in Figure \ref{fig:residual_slab_het} demonstrate the need for source iteration calculations between DMD updates. The DMD update does introduce some high-frequency errors that are quickly removed from the solution; these are apparent in the jumps in the residual after a DMD update.

\subsection{Multi-Dimensional Examples} 
\label{sec:two-d}
\begin{figure}[!htb]
  \centering
  \includegraphics[width=\textwidth]{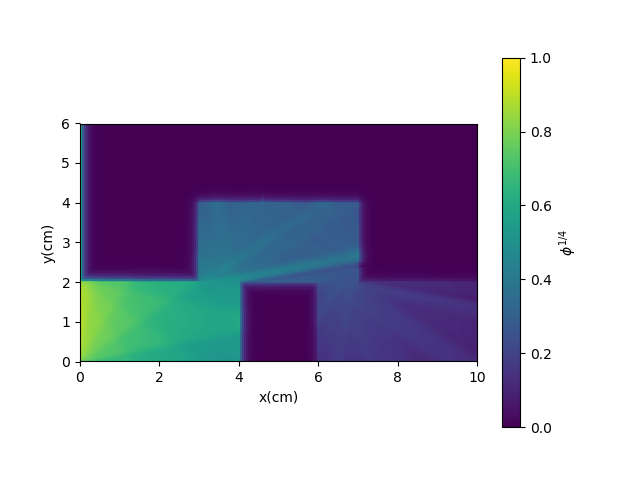}
  \caption{Crooked Pipe results with $S_8$.}   
  \label{fig:crooked_pipe}
\end{figure}

{A version of the crooked pipe problem \cite{graziani2000crooked} is a more realistic test.}
We solve a steady, linear, xy-geometry version of the crooked pipe problem where all materials have a scattering ratio of $0.988$ (to simulate the time-absorption of a realistic sized time step).
 The density ratio between the thick and thin material is 1000.
 We solve the problem using fully lumped, bilinear discontinous Galerkin in space and $S_8$ product quadrature.  The solution using a $200 \times 120$ grid of cells for the domain of size $10 \times 6$ mean-free paths is shown in Figure \ref{fig:crooked_pipe}. For this problem with different mesh resolutions we observe slow growth of the number of transport sweeps needed for the solution, this is not present in the source iteration calculation.  The number of transport sweeps to complete the solve with $K = 10$ and $R = 3$ is shown in Table \ref{table:cp}.
The increase seems to be the resolution to the 1/2 power (square root of the number of cells per dimension).

\begin{table}[!htb]
  \centering
  \caption{\bf Number of transport sweeps to solve the crooked pipe problem. Source iteration was not performed for the high-resolution calculations due to the excessive cost.}\label{table:cp}\begin{tabular}{r|rr}
\hline
  $(N_x \times N_y)$ &  DMD & SI\\
\hline
  $25 \times 15$ & 53 & 811 \\
  $50 \times 25$ & 52 & 873 \\
  $100 \times 60$ & 78 & 974 \\
  $150 \times 90$ & 91 & - \\
  $200 \times 120$ & 104 & - \\
\hline
\end{tabular}
\end{table}

\section{CONCLUSIONS}
Our results demonstrate the DMD can be used to accelerate the source iteration procedure, typically decreasing the number of transport sweeps required by an order of magnitude.  We believe this method will be valuable for situations where the effectiveness of standard DSA preconditioning is not assured. This includes high-order meshes, schemes with negativity fixes, and unstructured meshes with cycles.

{There are opportunities to this approach beyond the acceleration strategy outlined above.} We could use DMD acceleration to compute a low-order transport acceleration (the so-called TSA method). In this case the we would use low-order in angle transport sweeps to estimate the slowly converging modes.
 Additionally, it is possible to estimate $\tilde{A}$ using independently generated vectors. This would enable the $Y_\pm$ matrices to be generated using sweeps computed in parallel.
 The big win could be from applying this to other iterative components such as energy group iterations or temperature iterations in radiative transfer.



\section*{ACKNOWLEDGEMENTS}
 This work was performed under the auspices of the U.S. Department of Energy by Lawrence Livermore National Laboratory under Contract DE-AC52-07NA27344. LLNL-ABS-764003-DRAFT.
 
 Work by RGM was supported by Lawrence Livermore National Laboratory under research subcontract B627130. 



\setlength{\baselineskip}{12pt}
\bibliographystyle{unsrt}
\bibliography{radtran}


{
Disclaimer: This document was prepared as an account of work sponsored by an agency of the United States government. Neither the United States government nor Lawrence Livermore National Security, LLC, nor any of their employees makes any warranty, expressed or implied, or assumes any legal liability or responsibility for the accuracy, completeness, or usefulness of any information, apparatus, product, or process disclosed, or represents that its use would not infringe privately owned rights. Reference herein to any specific commercial product, process, or service by trade name, trademark, manufacturer, or otherwise does not necessarily constitute or imply its endorsement, recommendation, or favoring by the United States government or Lawrence Livermore National Security, LLC. The views and opinions of authors expressed herein do not necessarily state or reflect those of the United States government or Lawrence Livermore National Security, LLC, and shall not be used for advertising or product endorsement purposes. }

\end{document}